\documentclass[12pt]{iopart}
\usepackage{epsf}

\begin{document}

\title[]{Multi-strange particle production in relativistic heavy ion collisions at $\sqrt{s_{NN}}=62.4$ GeV}
\author{G. M. S. Vasconcelos ({\it for the STAR Collaboration})}
\address{Instituto de F\'{i}sica Gleb Wataghin, Universidade Estadual de Campinas - UNICAMP, Brazil}
\ead{gmagela@ifi.unicamp.br}

\begin{abstract}
We present preliminary STAR results on measurements of multi-strange particles $\Xi$, $\Omega$ and their anti-particles from Au+Au and Cu+Cu at $\sqrt{s_{NN}}=62.4$ GeV collisions. In order to better understand the role of strangeness enhancement in nucleus-nucleus collisions and its scaling properties with system size, we compare the results from Au+Au and Cu+Cu reactions for different event centrality classes. Strangeness enhancement is discussed in the context of multi-strange to pion ratios. Finally, $\Omega/\phi$ ratio is shown for different systems and energies for a systematic study. 
\end{abstract}

\pacs{25.75.-q, 25.75.Dw}

\section{Introduction}

For more than 20 years strangeness enhancement has been proposed as one of the signatures of the Quark Gluon Plasma (QGP) formation~\cite{rafelski}. Currently there are many experimental results from SPS and RHIC showing the enhancement of strange particles, such as $\Lambda$, $\Xi$, $\Omega$ and $\phi$ from different energies in A+A collisions when compared to p+p results~\cite{NA57,seSTAR1,seSTAR2,phiSTAR}.

The intermediate transverse momentum range ($2<p_{T}<5$ GeV/$c$) plays an important role in understanding hadronisation mechanisms. In this $p_{T}$ range, an anomalous increase of the $p/\pi$ ratio was observed in heavy ion collisions compared to p+p collisions in the RHIC data~\cite{ppiSTAR1,ppiSTAR2}. The enhancement of the baryon to meson ratio at intermediate transverse momentum has been also confirmed in the strangeness sector using ($\Lambda/K^{0}_{s}$)~\cite{lamk0Anthony}. In this context, the $\Omega/\phi$ ratio is an important measurement because it involves only strange quarks~\cite{omegaphicoa}. One of the theoretical models that try to explain the enhancement of baryon to meson ratio is the coalescence model, which assumes that hadrons are formed by free quarks~\cite{coalescence}. Calculation from Hwa and Yang~\cite{HwaYang} shows that the production of $\Omega$ and $\phi$ arises mainly from the recombination of thermal partons, thus exhibiting exponential $p_{T}$ dependence. Baryons at a certain $p_{T}$ come from partons of a lower average $p_{T}$ ($\sim p_{T}/3$) than mesons (formed by partons with $\sim p_{T}/2$). 

In this proceedings, we show recent results of multi-strange baryons for Au+Au and Cu+Cu at $\sqrt{s_{NN}} = 62.4$ GeV at mid-rapidity ($|y|<1$). With their high mass and strange quark content, $\Xi$ and $\Omega$ are important measurements of strange particle production and the formation of those should be less influenced by the net-baryon density. A comparison between Au+Au and Cu+Cu allows to probe the system size dependence, and the collisions at $\sqrt{s_{NN}} = 62.4$ GeV provide an important connection between the energy available at SPS ($\sqrt{s_{NN}} = 17.3$ GeV) and the top RHIC energy ($\sqrt{s_{NN}} = 200$ GeV).  

\section{Data analysis}

To perform the analysis presented here, we used minimum-bias Au+Au and Cu+Cu collision events at $\sqrt{s_{NN}} = 62.4$ GeV taken by the STAR experiment during run 2004 ($\sim$7 $M$ events from Au+Au) and run 2005 ($\sim$9 $M$ events from Cu+Cu).

In the STAR detectors, charged particle tracks were reconstructed by a large cylindrical Time Projection Chamber (TPC) and the ionization energy loss in the TPC gas was used for identification of charged particles. Strange particles were reconstructed using their weak decay topology: $\Lambda \rightarrow p+\pi^{-}$ (branching ratio of 63.9$\%$~\cite{pdg}); $\Xi^{-} \rightarrow \Lambda + \pi^{-}$ (branching ratio of 99.887$\%$~\cite{pdg}) and $\Omega^{-} \rightarrow \Lambda + K^{-}$ (branching ratio of 67.8$\%$~\cite{pdg}). The association of two or three daughter particles leads to a large amount of combinatorial background. To reduce the background, some geometrical cuts were applied to obtain a significant signal. However, some background remained underneath the mass peak and was therefore subtracted by making an interpolation of the spectrum on either side of the mass peak. Detector acceptance, tracking and selection inefficiencies were corrected by factors estimated using Monte Carlo generated particles that were propagated through a full GEANT~\cite{geant} simulation of the STAR detector and were added into real events. Then, this embedded event was processed with the standard event reconstruction chain.

\section{Results}

\begin{figure}
\centerline{\epsfxsize 3.1in \epsffile{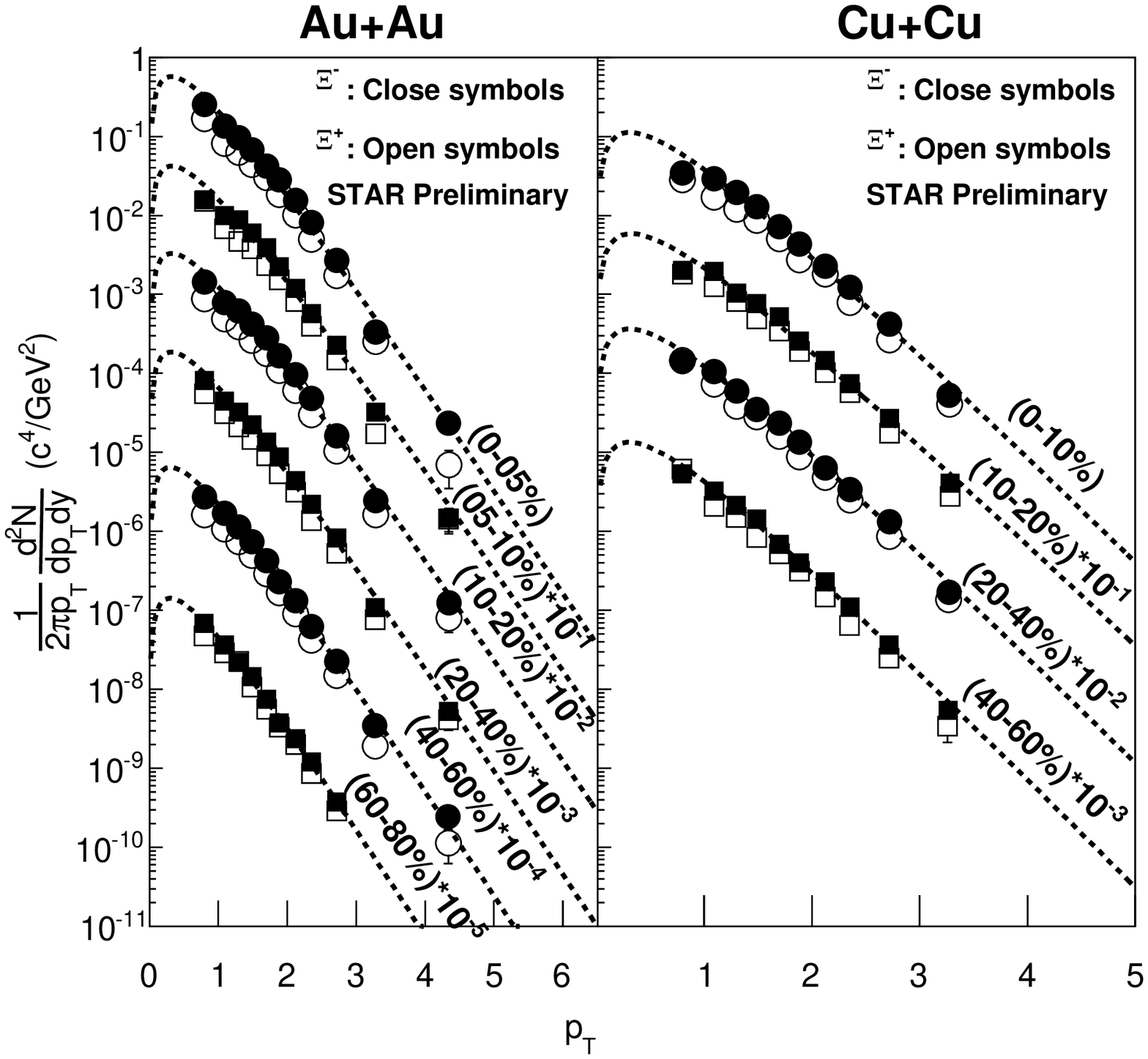} \epsfxsize 3.1in \epsffile{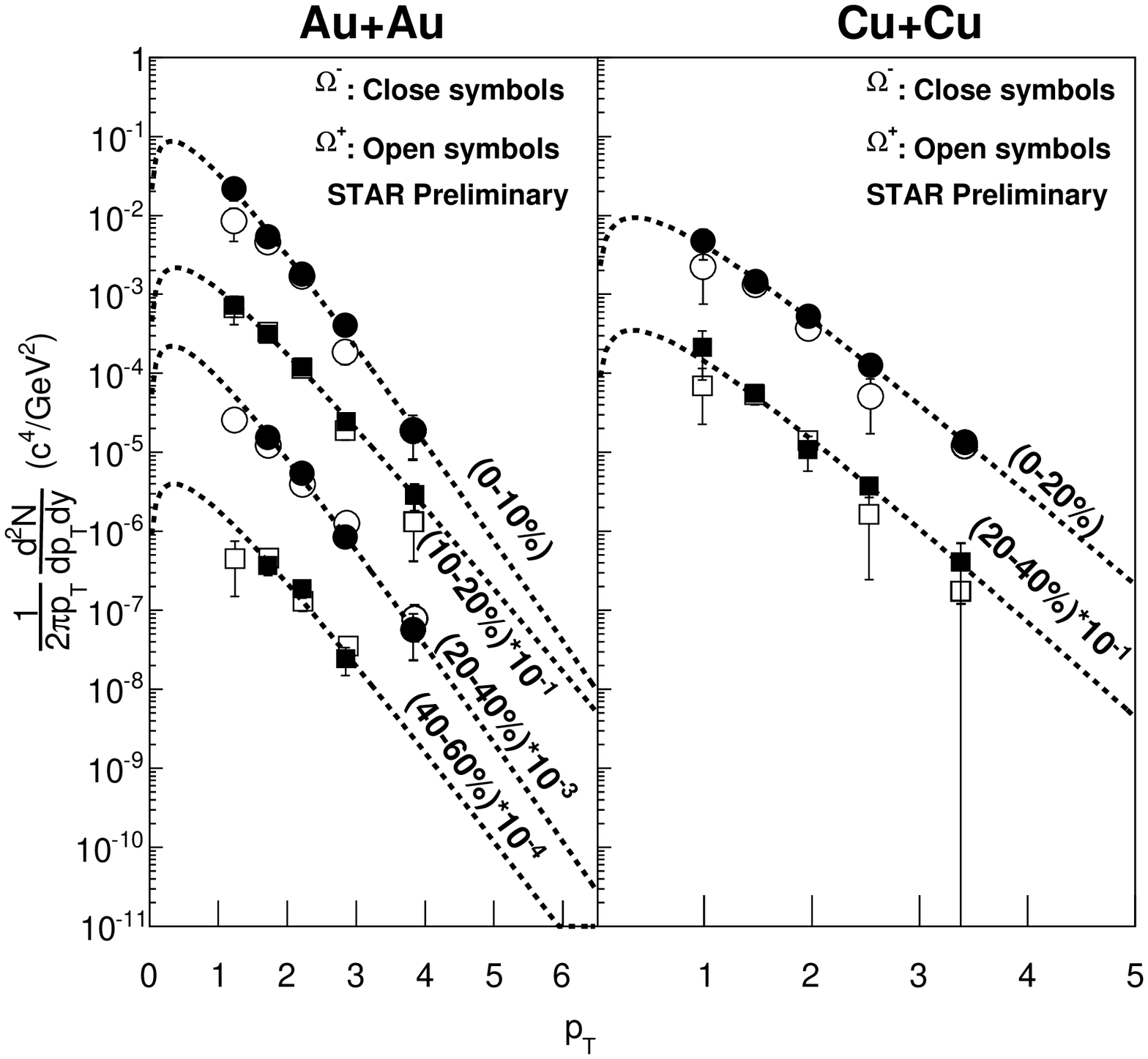}}
\caption{\label{figure1} Transverse momentum spectra for $\Xi^{-}$($\Xi^{+}$) on the left panel and $\Omega^{-}$ ($\Omega^{+}$) on the right for Au+Au and Cu+Cu at $\sqrt{s_{NN}} = 62.4$ GeV. For better visualization, the spectra were divided by factors of 10 and the centralities are listed in table~\ref{tab1}. Errors are statistical only.}
\end{figure}

Transverse momentum distributions of $\Xi$, $\Omega$ and their anti-particles at mid-rapidity ($|y|<$1) are shown in figure~\ref{figure1} for different centrality classes. The Boltzmann function was used to fit those spectra in the low $p_{T}$ region allowing the determination of the integrated yield ($dN/dy$) as well as the inverse slope parameter. We noted that the $dN/dy$ monotonically increases with centrality and table~\ref{tab1} summarizes the centralities used in this analysis as well as the yields and the inverse slope parameter.

\begin{table}
\centering
\caption{\footnotesize{Yields and inverse slope parameter (in MeV) for $\Xi^{-}$ and $\Omega^{-}$.}}
\small\rm
\begin{tabular}{c|c|c|c|c|c|c}
\hline
\hline
& \multicolumn{3}{c}{Au+Au} \vline & \multicolumn{3}{c}{Cu+Cu}\\
\hline
& Centrality & Inv. Slope & $dN/dy$ & Centrality & Inv. Slope & $dN/dy$ \\
\hline
$\Xi^{-}$ & 0 $-$ 5$\%$  & 336 $\pm$ 4 & 1.75  $\pm$ 0.07  & 0 $-$ 10$\%$  & 322 $\pm$ 3 & 0.32 $\pm$ 0.01   \\
         & 5 $-$ 10$\%$  & 339 $\pm$ 5 & 1.38  $\pm$ 0.06  &               &             &                   \\
         & 10 $-$ 20$\%$ & 336 $\pm$ 3 & 1.07  $\pm$ 0.03  & 10 $-$ 20$\%$ & 328 $\pm$ 4 & 0.188 $\pm$ 0.008 \\
         & 20 $-$ 40$\%$ & 329 $\pm$ 3 & 0.57  $\pm$ 0.01  & 20 $-$ 40$\%$ & 313 $\pm$ 4 & 0.107 $\pm$ 0.005 \\
         & 40 $-$ 60$\%$ & 313 $\pm$ 5 & 0.200 $\pm$ 0.008 & 40 $-$ 60$\%$ & 303 $\pm$ 7 & 0.039 $\pm$ 0.003 \\
         & 60 $-$ 80$\%$ & 298 $\pm$ 7 & 0.043 $\pm$ 0.003 &               &             & \\
\hline
$\Omega^{-}$ & 0 $-$ 10$\%$ & 347$\pm$ 26 & 0.26 $\pm$ 0.07   & 0 $-$ 20$\%$  & 353 $\pm$ 29 & 0.038 $\pm$ 0.008 \\
            & 10 $-$ 20$\%$ & 381$\pm$ 26 & 0.12 $\pm$ 0.02   &               &              &                   \\
            & 20 $-$ 40$\%$ & 349$\pm$ 19 & 0.06 $\pm$ 0.01   & 20 $-$ 40$\%$ & 317 $\pm$ 38 & 0.015 $\pm$ 0.006 \\
            & 40 $-$ 60$\%$ & 352$\pm$ 31 & 0.018 $\pm$ 0.005 &               &              &                   \\
\hline
\end{tabular}
\label{tab1}
\end{table}

\begin{figure}
\centerline{\epsfxsize 3.1in \epsffile{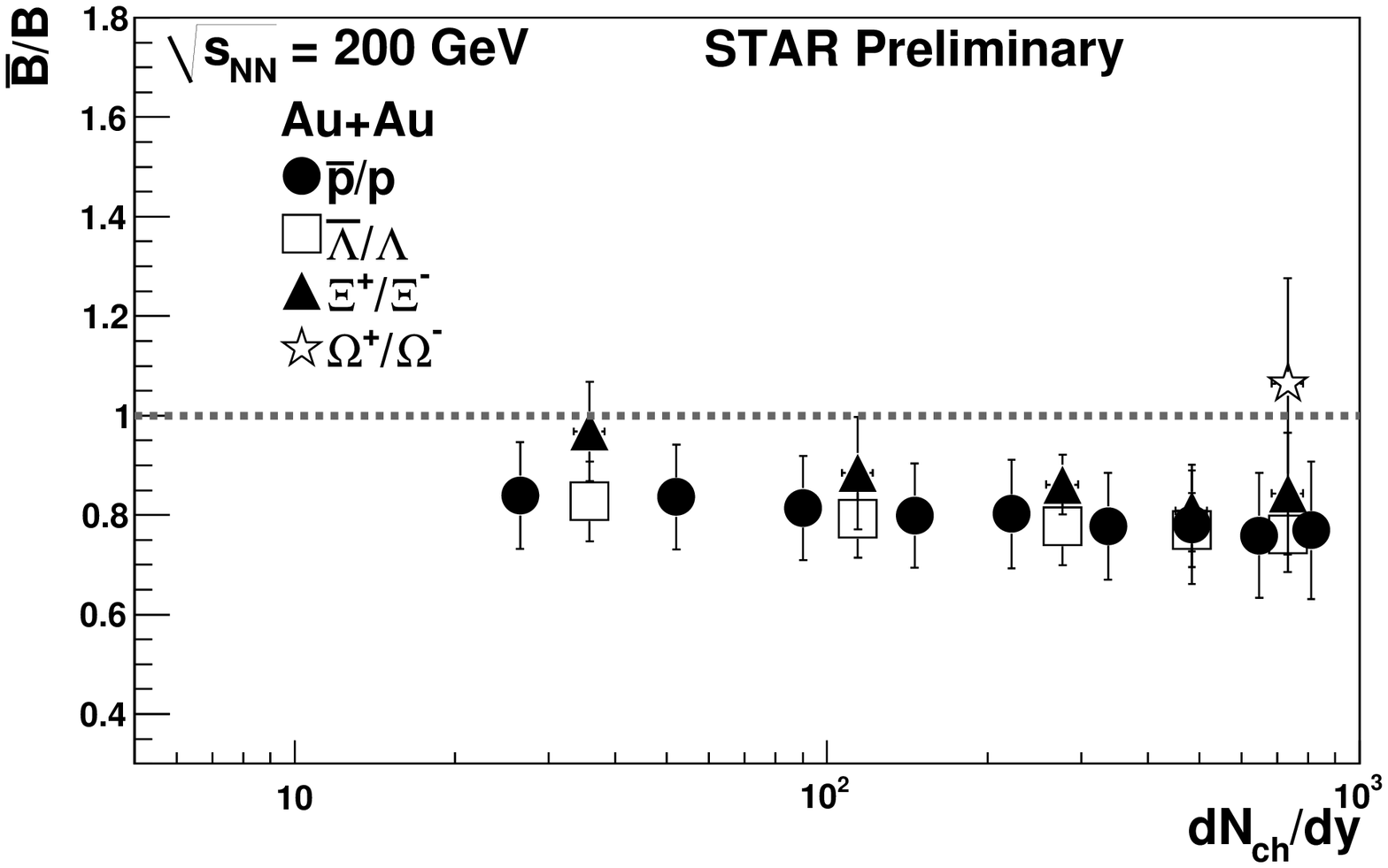} \epsfxsize 3.1in \epsffile{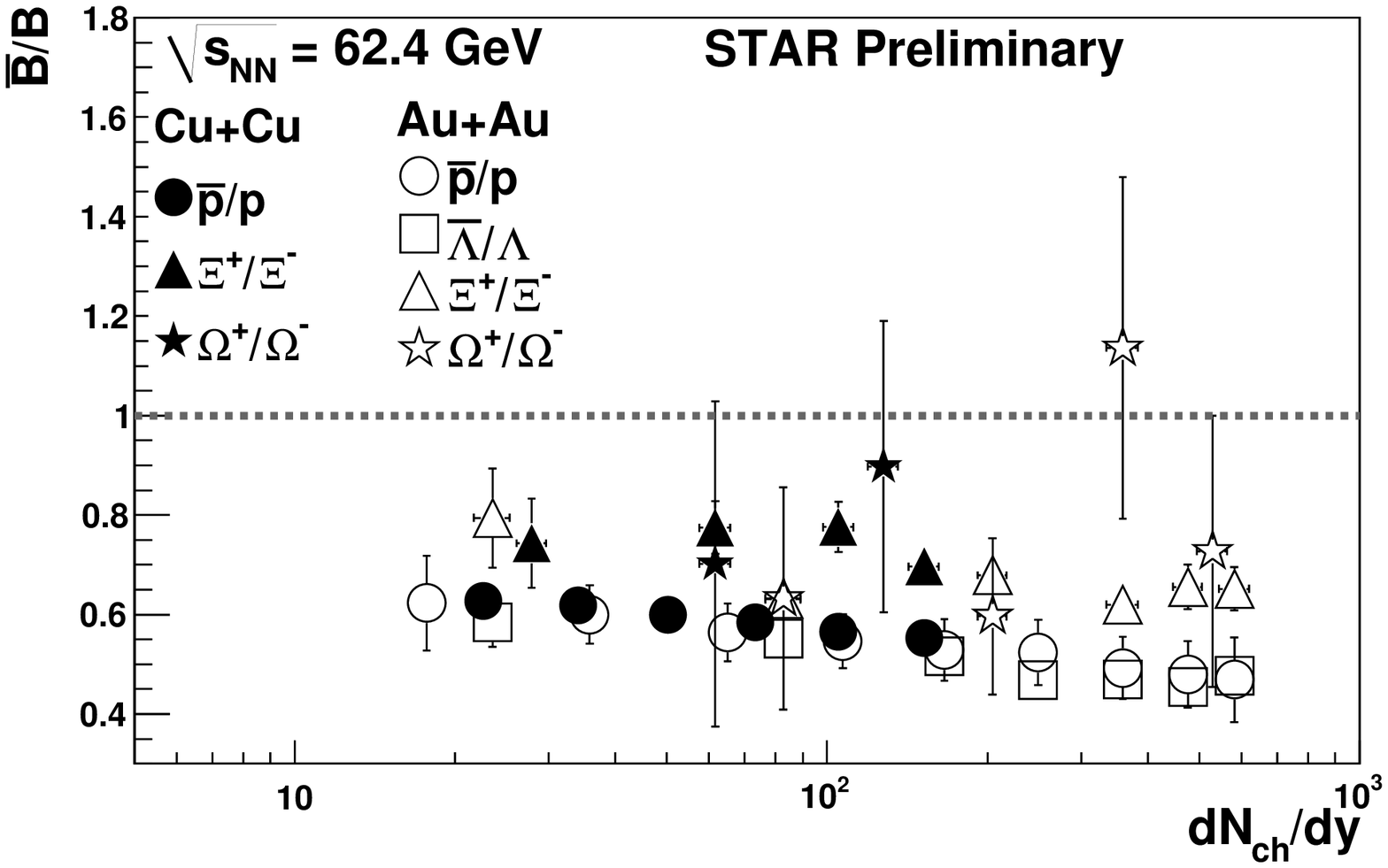}}
\caption{\label{figure2} Anti-baryon to baryon ratio as a function of $dN_{ch}/dy$ for collisions at $\sqrt{s_{NN}} = 200$ GeV on the left panel and at$\sqrt{s_{NN}} = 62.4$ GeV on the right panel.}
\end{figure}

Figure~\ref{figure2} shows the anti-baryon to baryon ratio as a function of charged particle multiplicity, $dN_{ch}/dy$ \footnote{Values of $dN_{ch}/dy $ were taken from~\cite{fuqiang}.}. The left panel on figure~\ref{figure2} shows those ratios for Au+Au at $\sqrt{s_{NN}} = 200$ GeV and it can be noted that all particle species have the same constant behavior as a function of centrality. A different behavior is observed at lower energy, as can be seen in the right panel, where the same ratios mentioned before are shown, but now for Au+Au and Cu+Cu at $\sqrt{s_{NN}} = 62.4$ GeV. For a given particle ratio, we observe no difference between Cu+Cu and Au+Au when using $dN_{ch}/dy$ as the scaling factor. The $\bar{p}/p$ and $\bar{\Lambda}/\Lambda$ ratios measured at $\sqrt{s_{NN}} = 62.4$ GeV decrease as the collisions become more central that could be an indication of a higher net-baryon density for central collisions. Multi-strange baryons have flat behavior for collision systems which shows they are less sensitive to changing net-baryon densities as a function of centrality.

\begin{figure}
\centerline{\epsfxsize 3.1in \epsffile{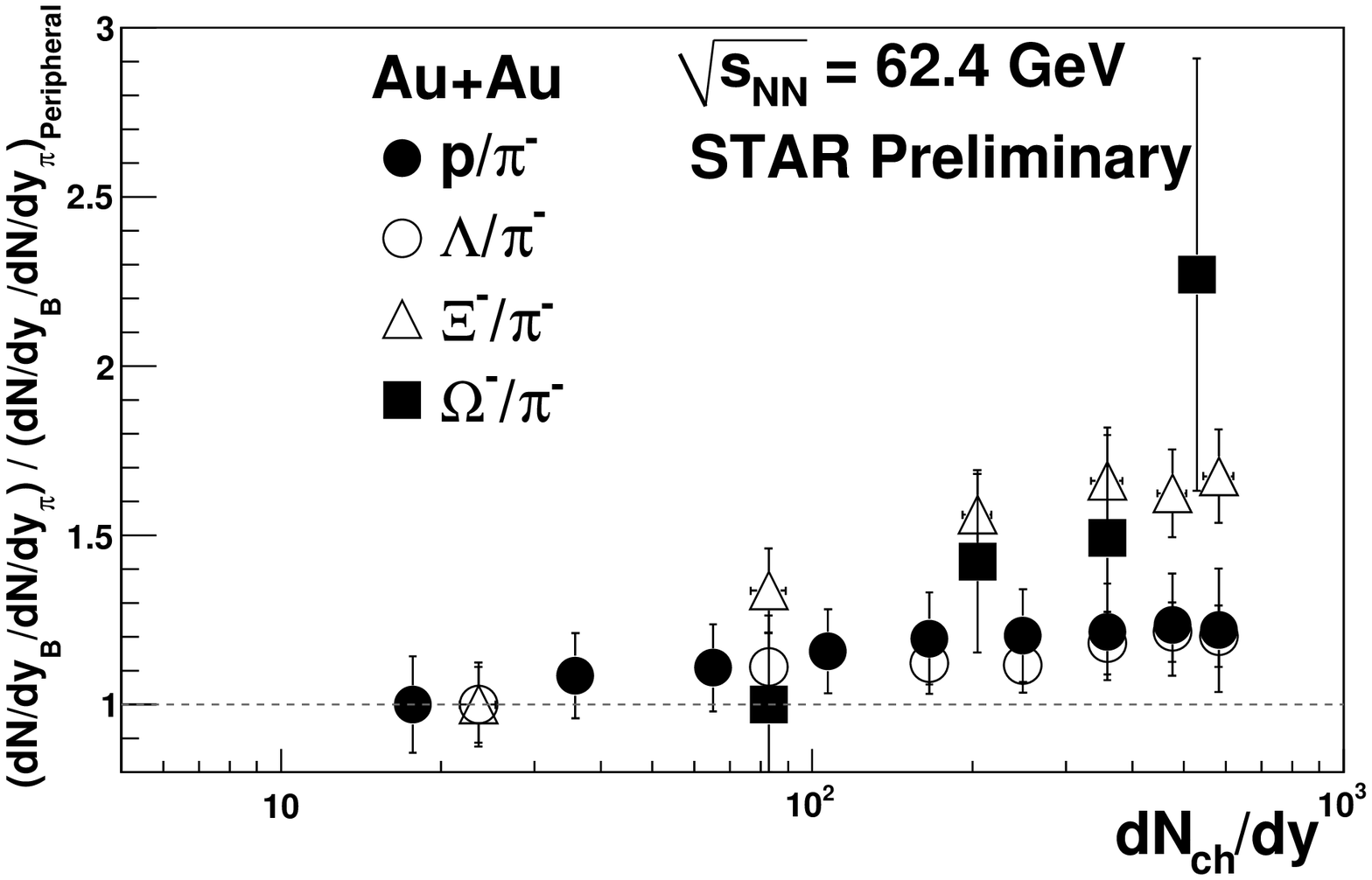} \epsfxsize 3.1in \epsffile{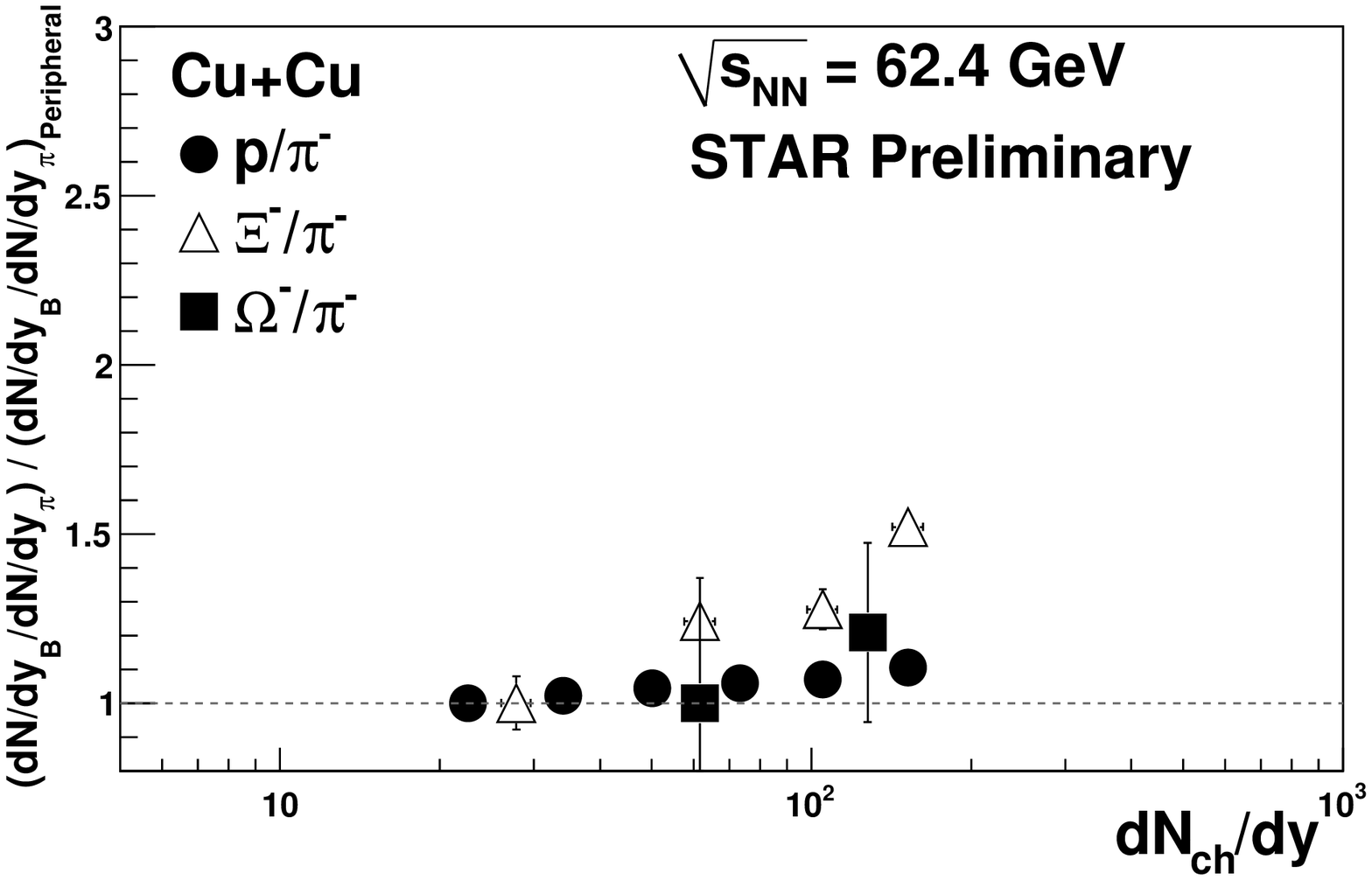}}
\caption{\label{figure3} Relative yield of baryon over $\pi^{-}$ yield normalized to the most peripheral data as function of $dN_{ch}/dy$ for Au+Au (left panel) and Cu+Cu (right panel) at $\sqrt{s_{NN}} = 62.4$ GeV.}
\end{figure}

Originally, the strangeness enhancement is defined as the ratio of strange particle yields from A+A collisions with p+p at the same energy. As we do not have results from p+p collisions at $\sqrt{s_{NN}} = 62.4$ GeV for $\Lambda$, $\Xi$ and $\Omega$, the $\pi^{-}$ yield was used as our reference. Figure~\ref{figure3} shows the ratio between baryons and charged pion ($\pi^{-}$) as a function of $dN_{ch}/dy$ normalized to the most peripheral centrality bin. As the proton contains no strange quark, the $p/\pi^{-}$ can be taken as a baseline for comparison. We observe that $\Lambda/\pi^{-}$ and $p/\pi^{-}$ present the same behavior and the same enhancement within uncertainties. Increasing the strange quark content we expect to observe a more pronounced enhancement which is confirmed by the multi-strange relative ratios showed. The $\Xi^{-}/\pi^{-}$ and $\Omega^{-}/\pi^{-}$ ratios have larger increase rates than other ratios in figure~\ref{figure3}. For Au+Au (left panel), the enhancement rate seems to follow the hierarchy of strange quark content as obtained in previous analysis for $\sqrt{s_{NN}} = 200$ GeV~\cite{seSTAR1}. The Cu+Cu data show the same trend compared to Au+Au data. In the Cu+Cu data it is difficult to conclude that there is still a hierarchy of the strangeness enhancement with the strange quark content.

\begin{figure}
\centerline{\epsfxsize 3.1in \epsffile{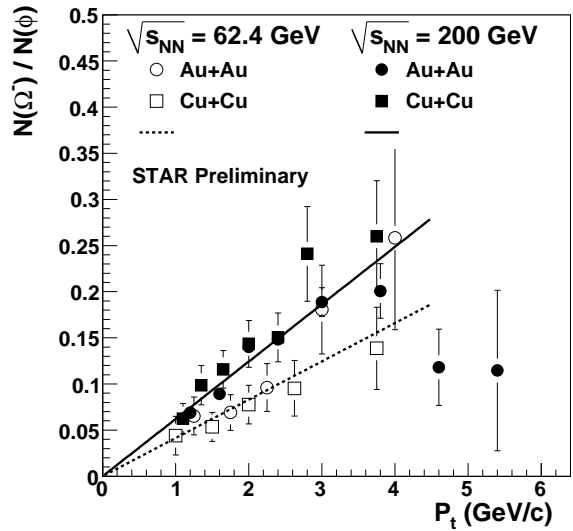}}
\caption{\label{figure4} $\Omega/\phi$ ratio as a function of transverse momentum for Au+Au and Cu+Cu at $\sqrt{s_{NN}} = 62.4$ GeV (open symbols) and $\sqrt{s_{NN}} = 200$ GeV (close symbols).}
\end{figure}

Finally we discuss the baryon over meson ratio. As mentioned in the introduction, an increase of baryon over meson ratio has been observed at intermediate $p_{T}$. Previous results from STAR at $\sqrt{s_{NN}} = 200$ GeV showed the $\Omega/\phi$ and a recombination model proposed by Hwa and Yang~\cite{HwaYang} described the enhancement of $\Omega/\phi$ ratio for $p_{T}$ up to 4 GeV/$c$ successfully~\cite{xiaobin}. Basically, this theoretical prediction showed a linear dependence on $p_{T}$. For high $p_{T}$ the baryon over meson ratio starts to decrease due to the contribution from jet fragmentation. Figure~\ref{figure4} presents the $\Omega/\phi$ ratio at $\sqrt{s_{NN}} = 62.4$ GeV and $\sqrt{s_{NN}} = 200$ GeV. The data still show the linear dependence on transverse momentum for $p_{T}$ smaller than 4 GeV/$c$ for data at lower energy. There is no statistical difference on absolute value for Au+Au and Cu+Cu collisions at the same energy, but there is a difference in the rate of increase of each energy where the rate of increase of $\sqrt{s_{NN}} = 200$ GeV is about 50$\%$ higher than in $\sqrt{s_{NN}} = 62.4$ GeV.

\section{Summary}

We have presented new results for multi-strange baryons production measured at Au+Au and Cu+Cu collsions at $\sqrt{s_{NN}} = 62.4$ GeV. The transverse momentum spectra of $\Xi$, $\Omega$ and their anti-particles were showed for different collision centrality classes at mid-rapidity. According to anti-baryon over baryon ratio, different trend was noted at $\sqrt{s_{NN}} = 62.4$ GeV compared to $\sqrt{s_{NN}} = 200$ GeV. The slightly decrease observed at $\bar{p}/p$ and $\bar{\Lambda}/\Lambda$ but not seen on multi-strange ratios as a function of $dN_{ch}/dy$ could be an indication of a higher net-baryon density for central collisions. For both colliding systems, the multi-strange particle yields relative to pion showed a higher increase rate with centrality which is consistent with the enhancement of the $s\bar{s}$ pair production at central collisions. Finally, a linear increase of $\Omega/\phi$ ratio for intermediate $p_{T}$ was observed and follows the same trend compared to previous results at $\sqrt{s_{NN}} = 200$ GeV.

\section*{Acknowledments} 
We wish to thank the Conselho Nacional de Desenvolvimento Cient\'ifico e Tecnol\'ogico, CNPq, Brazil for the support to participate in this conference. We thank the RHIC Operations Group and RCF at BNL, the NERSC Center at LBNL and the Open Science Grid consortium for providing resources and support. This work was supported in part by the Offices of NP and HEP within the U.S. DOE Office of Science, the U.S. NSF, the Sloan Foundation, the DFG cluster of excellence `Origin and Structure of the Universe', CNRS/IN2P3, STFC and EPSRC of the United Kingdom, FAPESP CNPq of Brazil, Ministry of Ed. and Sci. of the Russian Federation, NNSFC, CAS, MoST, and MoE of China, GA and MSMT of the Czech Republic, FOM and NWO of the Netherlands, DAE, DST, and CSIR of India, Polish Ministry of Sci. and Higher Ed., Korea Research Foundation, Ministry of Sci., Ed. and Sports of the Rep. Of Croatia, Russian Ministry of Sci. and Tech, and RosAtom of Russia.


\section*{References}
\begin{thebibliography}{10}
\bibitem{rafelski} J. Rafelski and B. M\"uller, {\it Phys. Rev. Lett.} \textbf{48} (1982) 1066.
\bibitem{NA57} F. Antinori {\it et al.} (NA57 Collaboration), {\it J. Phys. G} \textbf{32} (2006) 427. 
\bibitem{seSTAR1} B. I. Abelev {\it et al.} (STAR Collaboration), {\it Phys. Rev. C} \textbf{77} (2008) 044908.
\bibitem{seSTAR2} J. Adams {\it et al.} (STAR Collaboration), {\it Phys. Rev. Lett.} \textbf{98} (2007) 062301. 
\bibitem{phiSTAR} B. I. Abelev {\it et al.} (STAR Collaboration), {\it Phys. Lett. B} \textbf{673} (2009) 183.
\bibitem{ppiSTAR1} B. I. Abelev {\it et al.} (STAR Collaboration), {\it Phys. Lett. B} \textbf{655} (2007) 104.
\bibitem{ppiSTAR2} B. I. Abelev {\it et al.} (STAR Collaboration), {\it Phys. Rev. Lett.} \textbf{97} (2006) 152301.
\bibitem{lamk0Anthony} A. R. Timmins (STAR Collaboration), {\it Nucl. Phys. A} \textbf{830} (2009) 829c.
\bibitem{omegaphicoa} L. W. Chen and C. M. Ko, {\it Phys. Rev. C} \textbf{73} (2006) 044903.
\bibitem{coalescence} R. Fries, V. Greco and P. Sorensen, {\it Annu. Rev. Nucl. Part. Sci.} \textbf{58} (2008) 177.
\bibitem{HwaYang} R. C. Hwa and C. B. Yang, {\it Phys. Rev. C} \textbf{75} (2007) 054904
\bibitem{pdg} C. Amsler {\it et al.}, {\it Phys. Lett. B} \textbf{667} (2008) 1.
\bibitem{geant} http://wwwasd.web.cern.ch/wwwasd/geant/
\bibitem{fuqiang} B. I. Abelev {\it et al.} (STAR Collaboration), {\it Phys. Rev. C} \textbf{79} (2009) 034909.
\bibitem{xiaobin} X. Wang (STAR Collaboration), {\it J. Phys. G} \textbf{35} (2008) 104074.
\end{thebibliography}
\end{document}